\documentclass[aps,pre,twocolumn,superscriptaddress,amsmath,floatfix]{revtex4-1}
\usepackage{graphicx}
\usepackage{subfigure}
\usepackage{amssymb}
\usepackage{physics}
\usepackage{xcolor}
\usepackage[normalem]{ulem}
\newcommand{\und}[1]{_\mathsf{#1}}

\begin{document}

\title{Maximal power for heat engines: role of asymmetric interaction times}

\author{Pedro E. Harunari}
\email{pedroharunari@gmail.com}
\affiliation{Instituto de F\'isica da Universidade de S\~ao Paulo,  05508-090 S\~ao Paulo, SP, Brazil}

\author{Fernando S. Filho}
\email{ffilho@if.usp.br}
\affiliation{Instituto de F\'isica da Universidade de S\~ao Paulo,  05508-090 S\~ao Paulo, SP, Brazil}

\author{Carlos E. Fiore}
\email{fiore@if.usp.br}
\affiliation{Instituto de F\'isica da Universidade de S\~ao Paulo,  05508-090 S\~ao Paulo, SP, Brazil}

\author{Alexandre Rosas}
\email{arosas@fisica.ufpb.br}
\affiliation{Departamento de F\'isica, CCEN, Universidade Federal da Para\'iba, Caixa Postal 5008, 58059-900, Jo\~ao Pessoa, Brazil}

\date{\today}

\begin{abstract}
  The performance of endoreversible thermal machines operating at finite power constitutes
  one of the main challenges of nonequilibrium classical and quantum thermodynamics,
  engineering and others. We introduce the idea of adjusting the interaction time
    asymmetry in order to optimize the engine performance.  We consider one of the simplest thermal machines,
  composed of a quantum dot interacting sequentially with two different reservoirs of heat
  and particles.  Distinct optimization protocols are analyzed in the framework of stochastic
  thermodynamics. Results reveal that asymmetric interaction times play a
  fundamental role in enhancing the power output and that maximizations can provide an
    increase larger than 25\% the symmetric case. As an extra advantage, efficiencies at maximum
    power are slightly greater than the endoreversible Curzon-Ahlborn efficiency for a broad range
  of reservoir temperatures.
\end{abstract}


\maketitle

\section{\label{intro}Introduction}

The efficiency of any heat engine is bounded by Carnot efficiency $\eta\und{C}= 1-T\und{C}/T\und{H}$, with $T\und{C}$ and $T\und{H}$ being the cold and hot reservoir temperatures. It constitutes one of the main results of Thermodynamics and is one of the distinct formulations of the second law. Such ideal limit was introduced by Carnot in 1824 \citep{carnot, de2013equilibrium} and consists of a reversible machine composed by two isothermal and two adiabatic quasi-static strokes. Although it is a universal upper bond valid for all engines, irrespective of their designs, compositions, nature, whether classical \citep{verley2014unlikely, martinez2016brownian} or quantum \citep{Vinjanampathy,kosloff}, such (ideal) limit is impractical, not only due to imperfections of the machine construction, which increases the dissipation, but also because its achievement would demand the machine to operate in a full reversible way during infinitely large times, implying its operation at a null power (finite work divided by infinite time).

Thus, it is usually desirable to build thermal machines as efficient as possible operating at finite power outputs.  
One of the main findings for endoreversible thermal machines is the Curzon and Ahlborn efficiency \citep{curzon1975efficiency}, in which the efficiency at maximum power is given  by $\eta_\text{CA} = 1- \sqrt{T\und{C} / T\und{H}}$. Such a remarkable finding has also been derived in several distinct works (see e.g.\ \cite{novikov1958efficiency, van2005thermodynamic}) and despite not possessing  the same universal status of the Carnot efficiency, it provides a powerful guide about the operation of nonequilibrum engines under more realistic situations and sheds light on the construction and performance of small-scale engines (nanoscopic devices) working at maximum power regime from the tools of stochastic thermodynamics
\citep{verley2014unlikely, martinez2016brownian, schmiedl2007efficiency, seifert2012stochastic, esposito2009universality, cleuren2015universality, van2005thermodynamic, esposito2010quantum, seifert2011efficiency, izumida2012efficiency, golubeva2012efficiency, holubec2014exactly, bauer2016optimal, proesmans2016power, tu2008efficiency, ciliberto2017experiments, akasaki2020entropy, tome2015stochastic}. In this context, single-level quantum dots have been proposed as prototype machines, whose simplicity allows detailed investigation of their performances at maximum power \citep{den2018need, scovil1959three, josefsson2018quantum}.

Collisional models, e.g.\ a system interacting sequentially and repeatedly with distinct environments (instead of continuous  interaction with all the reservoirs), have been considered as a suitable description of engineered reservoirs \cite{PhysRevResearch.2.043016}.
Among the distinct situations for that, we mention the case of quantum systems, in which the reservoir is conveniently represented as a sequential collection of uncorrelated particles \cite{landif, molitor2020stroboscopic}. Additionally, the collisional approach attempts to provide realistic systems interacting only with small fractions of the environment or even those evolving under the
influence of distinct drivings over each member \cite{parrondo1,esposito11}. Particularly, many aspects of a stochastic pump in which a single-level quantum dot (QD) is connected sequentially and periodically to different reservoirs have been discussed lately for  symmetric
  interaction times~\citep{rosas1,rosas2,PhysRevE.98.052137}.


In this work, we introduce
  the idea of adjusting the interaction time asymmetry in order to optimize the
  engine performance. The present approach is rather different 
  from some findings~\cite{mondal2020exploring,cavina2020maximum,schmiedl2007efficiency, bauer2016optimal} exactly because we
explore this adjustment of the interactions time, that is,
the interaction time is the focus of our study.  
          Despite the simplicity of the system, its large applicability and richness allows the usage as heat engine, refrigerator, heater or accelerator and     hence  highlighting the importance of searching for
optimized protocols.
  As a main finding,  under suited situations, asymmetric interaction times play an important role for the enhancement of power output. Also, as an extra advantage, efficiencies become somewhat greater than the endoreversible Curzon-Ahlborn efficiency.

This paper is organized as follows: In Sec.~\ref{model} the model is presented and analytical results are provided. Sec.~\ref{thermod} is devoted to the main results concerning the efficiency in different regimes and optimized power outputs. Lastly, in Sec.~\ref{discuss} we draw the conclusions and perspectives.

\begin{figure}[ht]
\begin{center}
\includegraphics[width=.43\textwidth]{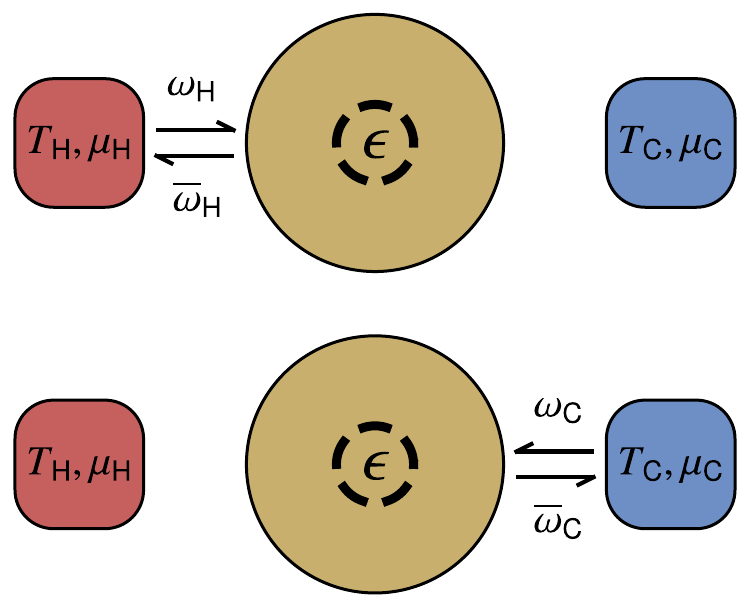}
\end{center}
\caption{Sketch of a quantum dot periodically and sequentially placed in contact with a hot (during a time \(\tau\und{H}\)) and a cold (during a time \(\tau - \tau\und{H}\)) reservoir. When in contact with a reservoir, the quantum dot receives a particle with rate \(\omega_i\) and donates a particle with rate \(\overline{\omega}_i\) (\(i\) being \(\mathsf{H}\) for the hot reservoir and \(\mathsf{C}\) for the cold one). When the quantum dot is occupied, its energy increases by \(\epsilon\).}
\end{figure}


\section{Model and exact solution\label{model}}

The model consists of a two-level system sequentially and periodically (with period $\tau$) placed in contact with a hot ($\mathsf{H}$) and a cold ($\mathsf{C}$) reservoir during the time intervals $\tau\und{H}$ and $\tau - \tau\und{H}$, respectively. More specifically, the QD interacts with one reservoir during a certain time. Afterwards, one turns off this interaction and then the QD is placed in contact with the second reservoir. The switching time is assumed to be instantaneous or, at least, much faster than any other relevant time scales. The energy of the QD is null ($\epsilon$) when it is empty (occupied by one electron). Each interaction can be modeled according to the transition rates $\omega_i$ and $\overline{\omega}_i$, whether
the system receives or delivers a particle, respectively, with $i \in \{\mathsf{H},\mathsf{C}\}$. Giving that the system placed in contact with a single reservoir evolves to the equilibrium distribution, the connection between transition rates and macroscopic quantities can be performed by assuming that the probability of occupation at equilibrium $p_i^\text{eq} \equiv \omega_i /(\omega_i+\overline{\omega}_i)$ obeys the Fermi-Dirac distribution $p_i^\text{eq} = [e^{(\epsilon -\mu_i)/T_i}+1]^{-1}$, where for each reservoir $T_i$ is the temperature, $\mu_i$ the chemical potential and the Boltzmann constant is set to $1$. The assumption above is equivalent to  the local detailed balance condition $\omega_i/\overline{\omega}_i = e^{-(\epsilon - \mu_i)/T_i}$ and, therefore, the temperature of each reservoir is given by 
\begin{equation}\label{temp}
  T_i = \frac{\mu_i - \epsilon}{\log \omega_i/\overline{\omega}_i}.
\end{equation}
As long as the reservoirs are different, the system will evolve to a periodic and time asymmetric non-equilibrium steady state (NESS), in which time-reversal means exchanging the order of the reservoirs. An important quantity is the ratio of the transition rates $\omega_i/ \overline{\omega}_i$, which  quantifies the reservoir willingness to concede a particle to the QD, equilibrium being reached for $\omega\und{H}/ \overline{\omega}\und{H} = \omega\und{C}/ \overline{\omega}\und{C}$.

We start the analysis of this system by considering a Markov chain whose discrete time is given by $t\equiv nh$, where $n=0,1,2,\ldots$ and $h$ is the time step. When the system is placed in contact with reservoir $i$, the transition matrix is given by
\begin{equation}\label{transmatrix}
	\mathcal{W}_i \equiv \begin{pmatrix}
	-\omega_i & \overline{\omega}_i \\
	\omega_i & -\overline{\omega}_i
	\end{pmatrix}.
\end{equation}
The probability distribution obeys the relation $\vec{P}_i(t+n h) = (\mathbb{I} +h \mathcal{W}_i)^n \vec{P}_i(t)$, where $\mathbb{I}$ is the $2\times2$ identity matrix and $\vec{P}_i(t)\equiv \{1-p_i(t),p_i(t)\}$ is the vector of probabilities of emptiness and occupation. Hence, $p_i(t+n h)$ can be written as 
\begin{equation}\label{markovprob}
	p_i(t+n h)= p_i^\text{eq}+ [1-h\overline{\omega}_i (1+\omega_i/ \overline{\omega}_i)]^n \left( p_i(t) - p_i^\text{eq} \right).
\end{equation}
Since $p_i(t)$ is continuous, one has the boundary conditions $p_\mathsf{H} (\tau\und{H}) = p_\mathsf{C} (\tau\und{H})$. Furthermore, the periodicity of the system ensures that it returns to the initial state after a complete period for long enough times, such that $p\und{H} (0) = p\und{C}(\tau)$. The occupation probability can be exactly obtained considering these boundary conditions and solving Eq. \eqref{markovprob}. Hereafter we consider the NESS regime, for which such boundary conditions are valid.

The NESS particle flux at a given time interval $h$ is given by $J_\mathsf{i}(nh) \equiv [p_\mathsf{i} ((n+1)h) - p_\mathsf{i}(nh)]/h$, which is positive whenever more particles leave the reservoir $i$ towards the QD on average, and negative otherwise. By averaging $J_\mathsf{i}(nh)$ over a full cycle we have that  $\overline{J}_\mathsf{H}= (1 /\tau) \sum_{n=0}^{\tau\und{H}/h-1} J_\mathsf{H}(nh)h$ and $\overline{J}_\mathsf{C} = (1/\tau) \sum_{m= \tau\und{H}/h}^{\tau/h-1} J_\mathsf{C}(mh)h$, for $i \in \{{\mathsf{H,C}}\}$. Since no electron accumulation in the QD is possible,
all particles leaving a given reservoir must go to the other one, such that $\overline{J}_\mathsf{H}+ \overline{J}_\mathsf{C} = 0$. By considering the master equation regime, $h\to 0$ and $n\to \infty$ with $nh=t$ held fixed, the above probabilities and currents in the NESS become
\begin{widetext}
\begin{equation}\label{ph}
	p_\mathsf{H}(t) = \frac{\omega\und{H}}{\omega\und{H} + \overline{\omega}\und{H}} - \frac{e^{- ( \omega\und{H} + \overline{\omega}\und{H})t} [1- e^{- ( \omega\und{C}+ \overline{\omega}\und{C})(\tau - \tau\und{H})} ]}{1- e^{%
	-(\omega\und{H} + \overline{\omega}\und{H})\tau\und{H} - ( \omega\und{C} +\overline{\omega}\und{C})(\tau - \tau\und{H})} }
	\frac{\omega\und{H}\overline{\omega}\und{C} - \overline{\omega}\und{H}\omega\und{C}}{(\omega\und{H} + \overline{\omega}\und{H})(\omega\und{C} + \overline{\omega}\und{C})},\ t=[ 0,\tau\und{H}]\ (\mathrm{mod}\ \tau),
\end{equation}
\begin{equation}\label{pc}
	p_\mathsf{C}(t) =\frac{\omega\und{C}}{\omega\und{C} + \overline{\omega}\und{C}} - \frac{e^{- (\omega\und{C}+ \overline{\omega}\und{C})(t - \tau\und{H})} [1- e^{- (\omega\und{H} + \overline{\omega}\und{H})\tau\und{H}} ]}{1- e^{- (\omega\und{H} + \overline{\omega}\und{H})\tau\und{H} - (\omega\und{C} + \overline{\omega}\und{C})(\tau - \tau\und{H})} }\frac{\overline{\omega}\und{H} \omega\und{C} - \omega\und{H}\overline{\omega}\und{C}}{(\omega\und{H} + \overline{\omega}\und{H})(\omega\und{C} + \overline{\omega}\und{C})},\ t=[ \tau\und{H},\tau ]\ (\mathrm{mod}\ \tau),
\end{equation}
\begin{equation}\label{jcontinuous}
	\overline{J}_\mathsf{H}= -\overline{J}_\mathsf{C}= \frac{1}{\tau} \frac{(1-e^{-(\omega\und{H} + \overline{\omega}\und{H}) \tau\und{H}}) (1- e^{-(\omega\und{C} + \overline{\omega}\und{C}) (\tau-\tau\und{H})}   )}{1- e^{-(\omega\und{H} + \overline{\omega}\und{H}) \tau\und{H} -(\omega\und{C} + \overline{\omega}\und{C}) (\tau-\tau\und{H}) }} 	\frac{\omega\und{H}\overline{\omega}\und{C} - \overline{\omega}\und{H}\omega\und{C}}{(\omega\und{H} + \overline{\omega}\und{H})(\omega\und{C} + \overline{\omega}\und{C})}.
\end{equation}
\end{widetext}
We pause to make a few comments: first,  such results recover the findings from Refs. \citep{rosas1, harunari2020exact} for symmetric interaction times ($\tau\und{H}=\tau/2$). Second, for both discrete and continuous cases, the hot reservoir is ``more willing'' to concede particles than the cold reservoir when $\omega\und{H}/ \overline{\omega}\und{H} > \omega\und{C}/ \overline{\omega}\und{C}$, implying that  $\overline{J}_\mathsf{H}> 0$ and  $\overline{J}_\mathsf{C}<0$, which is consistent with the fact that $\overline{J}_\mathsf{i}$ points from reservoir $i$ to the QD. Third, the period $\tau$ only gives the time scale of the model in the sense that rescaling the fluxes, transition rates and the time the system stays in contact with each reservoir by $\tau$ keeps Eq.~(\ref{jcontinuous}) unchanged; hence, from now on  $\tau$ is kept fixed and reads $\tau=1$. Fourth and last, the present system can operate as a heat engine, refrigerator, heater or accelerator, provided the parameters $\omega_i$ and $\overline{\omega}_i$ (or equivalently, $\epsilon$, $\mu_i$ and $T_i$) are conveniently adjusted. In this paper we shall address the heat engine regime, which is set by the conditions $\{\omega\und{C}/ \overline{\omega}\und{C}< \omega\und{H}/ \overline{\omega}\und{H} <1, \mu_\mathsf{H}< \mu\und{C} <\epsilon \}$, ensuring the positiveness of temperatures (with $T_\mathsf{H} > T_\mathsf{C}$), power output and the heat extracted from the hot reservoir.


\section{Thermodynamics}\label{thermod}
Once obtained the probability distribution and the suited heat engine regime in terms the of model parameters, we are now in position to obtain the thermodynamic properties (exchanged heat and work) and efficiency through the framework of stochastic thermodynamics \citep{seifert2012stochastic, esposito}. Remarkable quantities averaged over a complete cycle are the exchanged heat and the chemical work given by $\overline{\dot{Q}}_i \equiv (\epsilon - \mu_i) \overline{J}_i$ and $\overline{\dot{W}}_i^\text{chem} \equiv \mu_i \overline{J}_i$, respectively, for $i=\{\mathsf{C},\mathsf{H} \}$. It is worth mentioning that they obey the first law of thermodynamics, in such a way that $\overline{\dot{Q}}\und{C} +\overline{\dot{Q}}_\mathsf{H}+\overline{\dot{W}}\und{C}^\text{chem}+\overline{\dot{W}}\und{H}^\text{chem}=0$.
For the engine regime, the efficiency is typically a measure of ``what you get and what you give'', signed here by  the ratio between the power output  $\overline{P}\equiv -(\overline{\dot{W}}^\text{chem}_\mathsf{H} + \overline{\dot{W}}^\text{chem}_\mathsf{C})$ and
the heat received from the hot reservoir $\overline{\dot{Q}}_\mathsf{H}$, resulting in 
\begin{equation}\label{eff}
	\eta \equiv \frac{\overline{P}}{\overline{\dot{Q}}\und{H}} =  1- \frac{T\und{C} \ln\omega\und{C}/ \overline{\omega}\und{C}}{T\und{H} \ln\omega\und{H}/ \overline{\omega}\und{H}}.
\end{equation}
The relation above can also be expressed in terms of the macroscopic properties of the reservoirs and the QD 
\begin{equation}\label{effmacro}
	\eta = 1 - \frac{\mu\und{C} - \epsilon}{\mu\und{H} - \epsilon} = \frac{\mu\und{C} - \mu\und{H}}{\epsilon - \mu\und{H}},
\end{equation}
where we used Eq.~(\ref{temp}). We pause again to make some comments: First, the system will reach an equilibrium state
when $\omega\und{H}/ \overline{\omega}\und{H}= \omega\und{C}/ \overline{\omega}\und{C}$, consistent with the (maximum) Carnot efficiency.
Second, the engine regime stated above $ (\omega\und{C}/ \overline{\omega}\und{C}< \omega\und{H}/ \overline{\omega}\und{H} <1), \; (\mu_\mathsf{H}< \mu\und{C} <\epsilon)$ ensures the positiveness of  the temperatures (with $T_\mathsf{H} > T_\mathsf{C}$), power output and the heat delivered from the hot reservoir.
Additionally, it is also consistent with heat flowing from the QD to the cold reservoir: \(\overline{\dot{Q}}\und{C} < 0\). Third  and last, since the right-hand-side of Eq.~\eqref{effmacro} does not depend on $\tau\und{H}$ and $\tau$,  $\eta$ is independent of the protocol. Conversely, the power output $\overline{P}$ depends on $\tau\und{H}$ and $\tau$, in such a way that it can be conveniently adjusted, together with the other parameters  $\{\omega\und{H}, \overline{\omega}\und{H}, \omega\und{C}, \overline{\omega}\und{C}, T\und{H}, T\und{C}\}$, in order to optimize the extracted power. Here, we are concerned with the maximization of power under different physical setups.

In order to exploit distinct possibilities of optimizing the extracted power, the next sections will be devoted to its maximization with respect to the protocol asymmetry and its complete maximization (also taking into account  the transition rates). Since the extracted power increases monotonically with the ratio between temperatures, the analysis will be carried out for finite fixed ratios $T\und{C}/T\und{H}$.

\subsection{Best protocol}

\begin{figure*}[ht]
  \begin{center}
%
    \subfigure[$T\und{H}=0.4, \; T\und{C}=0.1, \; \omega\und{H} = 0.04, \; \bar{\omega}\und{H} =0.18, \; \omega\und{C} = 0.01$ and $\bar{\omega}\und{C}=0.55$.]{\includegraphics[angle = 0,width =3in]{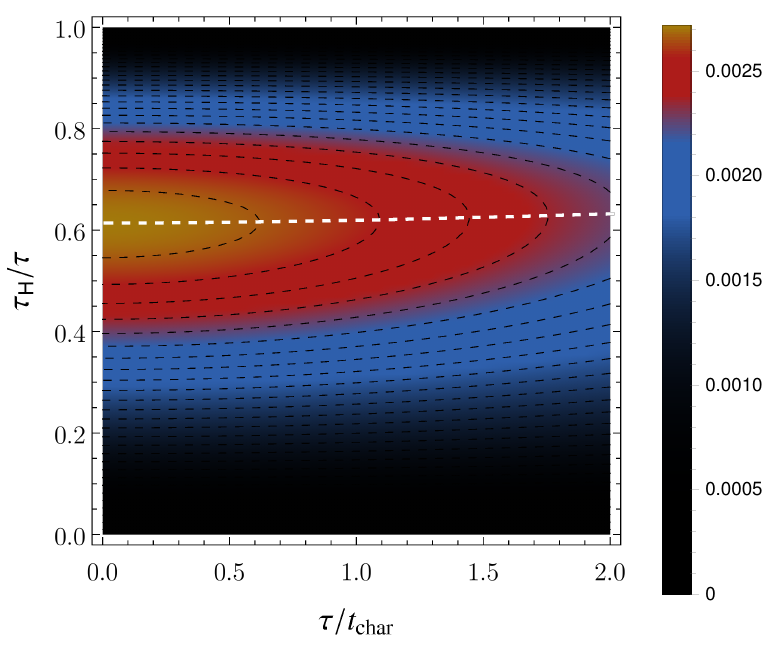}}
  \subfigure[$T\und{H}=2.0, \; T\und{C}=0.5, \; \omega\und{H} = 0.64, \; \bar{\omega}\und{H} = 0.86, \; \omega\und{C} = 0.06$ and $\bar{\omega}\und{C}=0.13$.]{\includegraphics[angle = 0,width =3in]{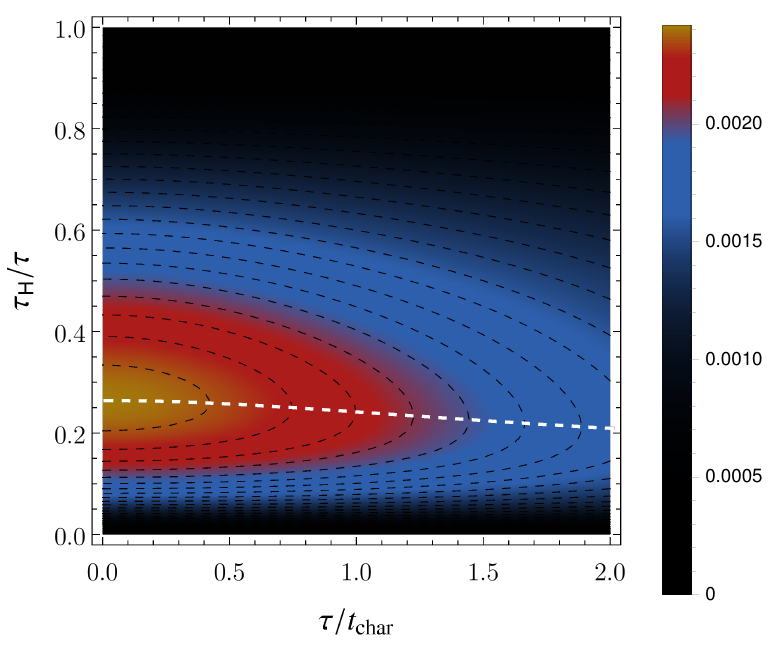}}\\
    \subfigure[$T\und{H}=3.5, \; T\und{C}=1.2, \; \omega\und{H} = 0.09, \; \bar{\omega}\und{H} = 0.11, \; \omega\und{C} = 0.71$ and $\bar{\omega}\und{C}=0.99$.]{\includegraphics[angle = 0,width =3in]{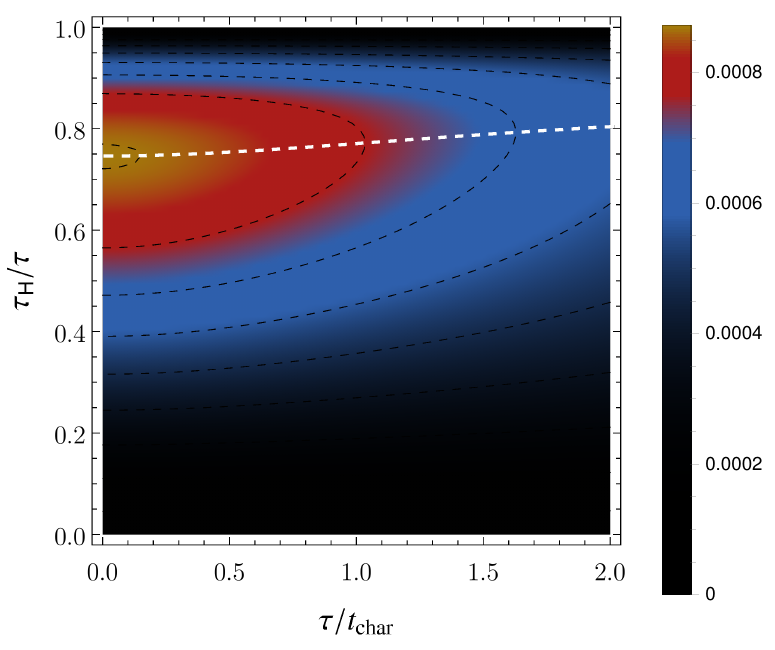}}
  \subfigure[$T\und{H}=0.4, \; T\und{C}=0.1, \; \omega\und{H} = 0.13, \; \bar{\omega}\und{H} = 0.58, \; \omega\und{C} = 0.01$ and $\bar{\omega}\und{C}=0.55$.]{\includegraphics[angle = 0,width =3in]{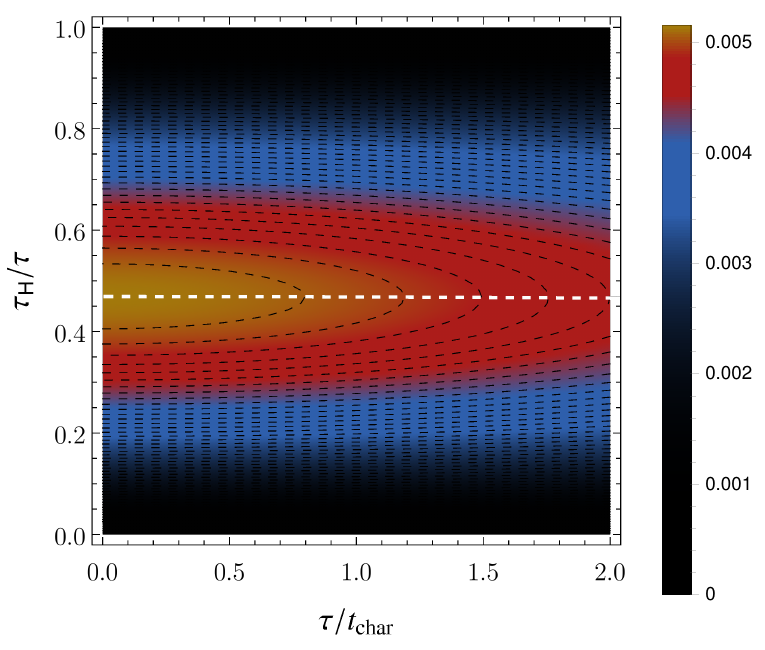}}\\
  \end{center}
  \caption{Power outputs versus $\tau/t_\text{char}$ and $\tau\und{H}/\tau$
    for distinct control parameters.
    Each dashed black curve represents an increment of $10^{-4}$ in the value of the power and the white dashed curve represents the value of $\tau\und{H}/\tau$ that maximizes the power for each value of $\tau/t_\text{char}$. For all panels, we set $\mu\und{H}=0.4$ and $\mu\und{C}=0.6$ and energy $\epsilon =1$.}
    \label{densities}
\end{figure*}

An interesting way of obtaining some insight about the system is to look at the power output as a function of the fraction of time spent in contact with the hot reservoir $\tau\und{H}$ and the ratio between the duration of the cycle and the characteristic time $t_\text{char}$ -- defined as the largest of characteristic times $t_\text{char}^\mathsf{H} \equiv  1/\left( \omega\und{H} + \bar{\omega}\und{H} \right)$ and $t_\text{char}^\mathsf{C} \equiv 1/\left( \omega\und{C} + \bar{\omega}\und{C} \right)$, which represent the typical relaxation time to the thermal state that each reservoir imposes -- as exemplified in Fig.~\ref{densities}, for some representative parameter values. In all cases, the chemical potentials and the energy are held fixed providing $\eta=1/3$ (see Eq.~(\ref{effmacro}) and caption of Fig.~\ref{densities} for the values used). A common trait of all panels is that the power output (and also the extracted heat) vanishes for $\tau\und{H}/\tau$ near 1 or 0. This is expected since in such situation the QD is mostly in contact with a single reservoir and almost no work can be extracted. It is also noteworthy that the value of $\tau\und{H}$ for which the power is maximum barely changes with the ratio $\tau/t_\text{char}$ but its value is extremely dependent on the other parameters of the model. Further, even for a constant value of $\tau/t_\text{char}$, the power output can be noticeably different.


In order to the reservoir effectively act upon the QD state it needs to interact during a period comparable with its characteristic time.
Therefore, characteristic times provide a notion for how long a good engine will interact with each reservoir -- e.g.\ if \(t_\text{char}^\mathsf{H} \ll t_\text{char}^\mathsf{C}\) the hot reservoir needs to interact very briefly compared to the cold one, so we can expect that a good protocol will present a small \(\tau\und{H} /\tau\). Hence, it is reasonable that good performance protocols will present interaction times around the characteristic times of each reservoir. This is exemplified in Fig.~\ref{densities}, where \(t_\text{char}^\mathsf{H} / ( t_\text{char}^\mathsf{H} + t_\text{char}^\mathsf{C} )\) gives approximately 0.7, 0.1, 0.9 and 0.4 for panels $a$, $b$, $c$ and $d$, respectively, in fair agreement with the best value of \(\tau\und{H}/\tau\) observed in the plots. As a rule of thumb we should choose $\tau$ as small as possible, in agreement with \cite{erdman2019maximum, mukherjee2020anti}, and maximize the power output with respect to $\tau\und{H}$. Although small, it is worth mentioning that the period $\tau$ should be sufficiently larger than the time necessary for switching the interaction of the QD from one reservoir to the latter one, in similarity with the symmetric case \citep{rosas1,PhysRevE.98.052137,PhysRevE.98.052137}.

The unique value of $\tau\und{H}$ that maximizes the $\overline{P}$ for a given value of $\tau$ and the other parameters can be easily found by numerically solving the transcendental equation
\begin{equation}
    \frac{(\omega\und{H}+\overline{\omega}\und{H})} {(\omega\und{C}+\overline{\omega}\und{C})} \frac{e^{- (\omega\und{H} + \overline{\omega}\und{H}) \tau\und{H}} [ 1- e^{-(\omega\und{C} + \overline{\omega}\und{C})  (\tau-\tau\und{H} )}]^2}{e^{- (\omega\und{C} + \overline{\omega}\und{C})  (\tau-\tau\und{H})} [ 1- e^{-(\omega\und{H} + \overline{\omega}\und{H})  \tau\und{H} }]^2} =1,
\end{equation}
whose values are shown in Fig.~\ref{densities}, represented by the white dashed lines.

\begin{figure*}[t]
    \centering
    \subfigure[$T\und{H}=2.0, \; \bar{\omega}\und{H} = 0.86$ and $\bar{\omega}\und{C}=0.14$.\label{densitiestratio_a}]{\includegraphics[angle = 0,width =3in]{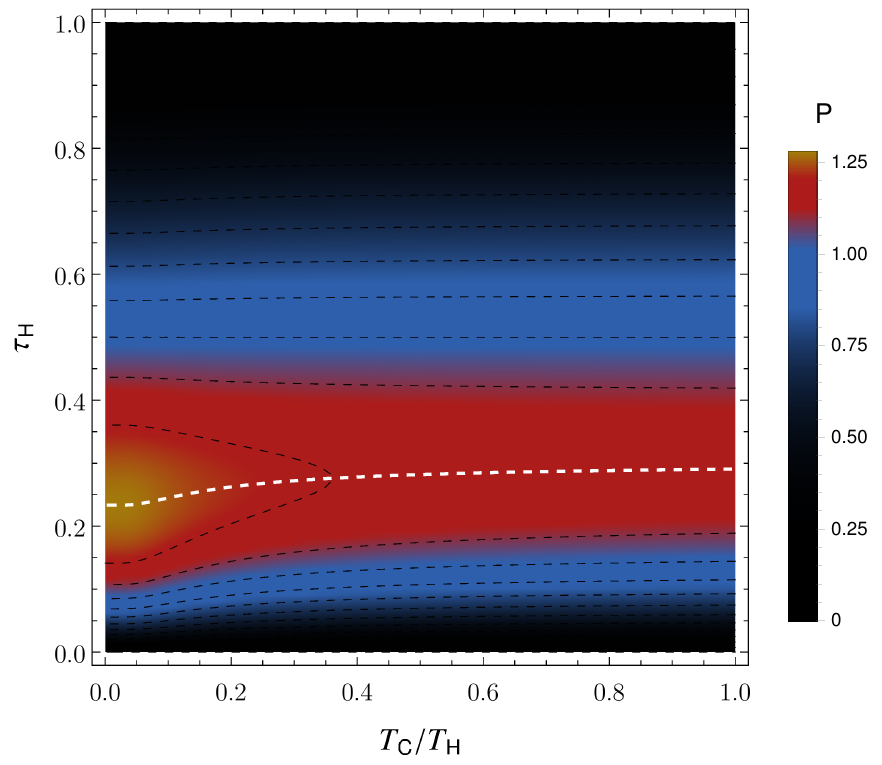}}
  \subfigure[$T\und{H}=0.4, \; \bar{\omega}\und{H} = 0.16$ and $\bar{\omega}\und{C}=0.69$.\label{densitiestratio_b}]{\includegraphics[angle = 0,width =3in]{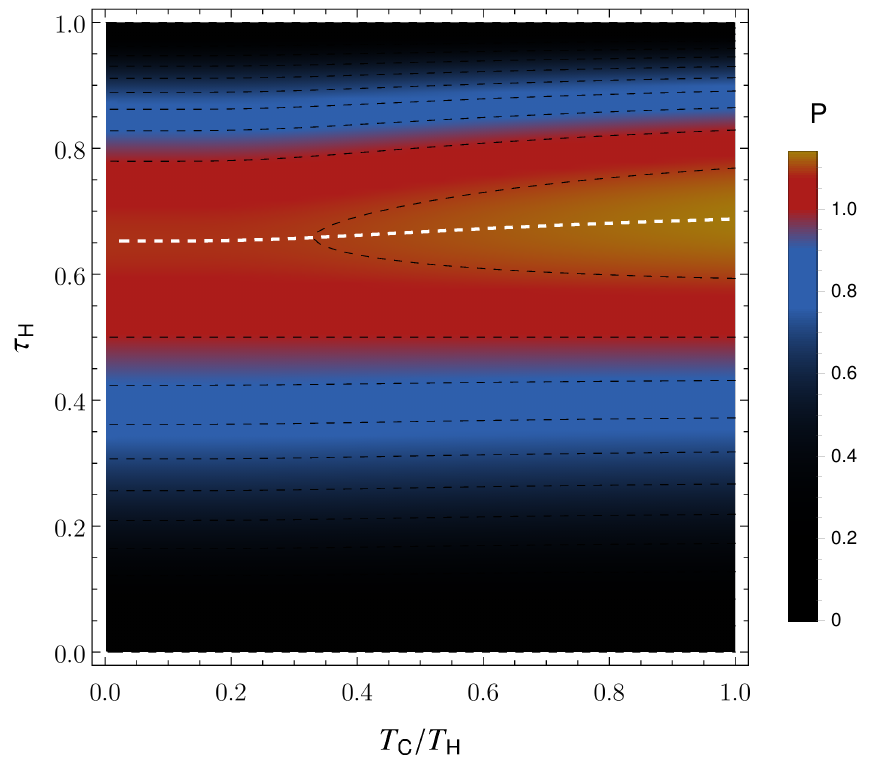}}\\
    \subfigure[$T\und{H}=2.0, \; \bar{\omega}\und{H} = 0.5$ and $\bar{\omega}\und{C}=0.5$.\label{densitiestratio_c}]{\includegraphics[angle = 0,width =3in]{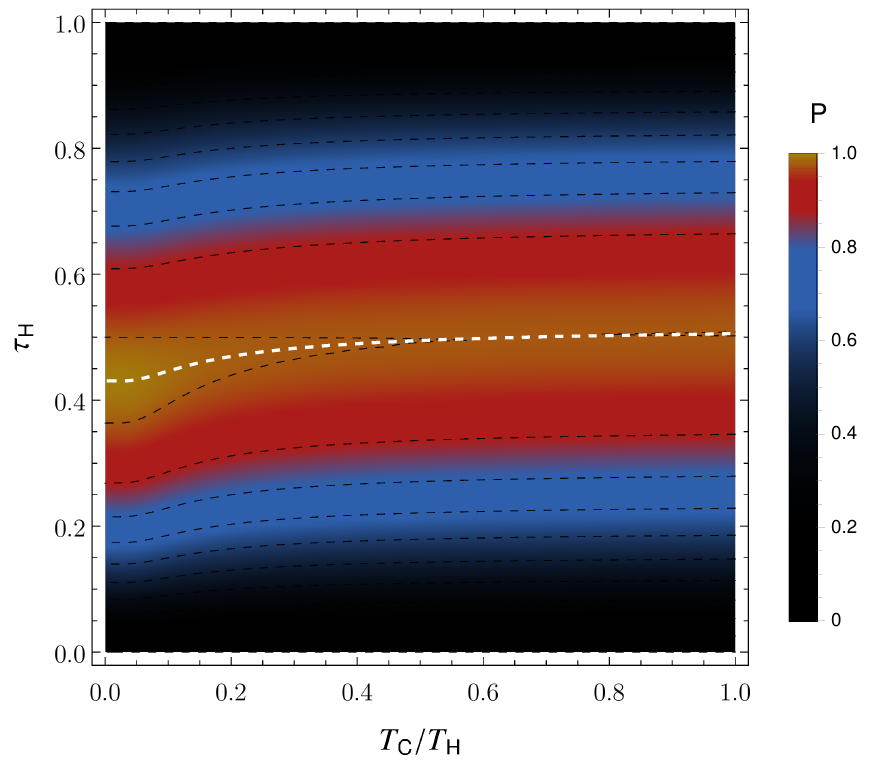}}
  \subfigure[$T\und{H}=0.4, \; \bar{\omega}\und{H} = 0.5$ and $\bar{\omega}\und{C}=0.5$.\label{densitiestratio_d}]{\includegraphics[angle = 0,width =3in]{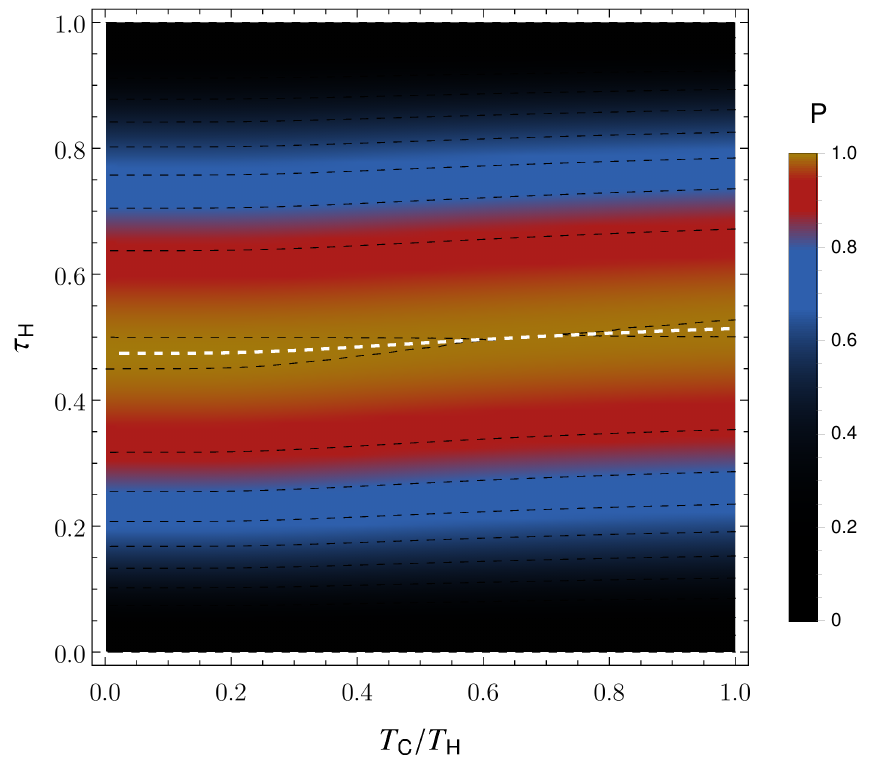}}\\
    \subfigure[$T\und{H}=0.3, \; \bar{\omega}\und{H} = 0.86$ and $\bar{\omega}\und{C}=0.14$.\label{densitiestratio_e}]{\includegraphics[angle = 0,width =3in]{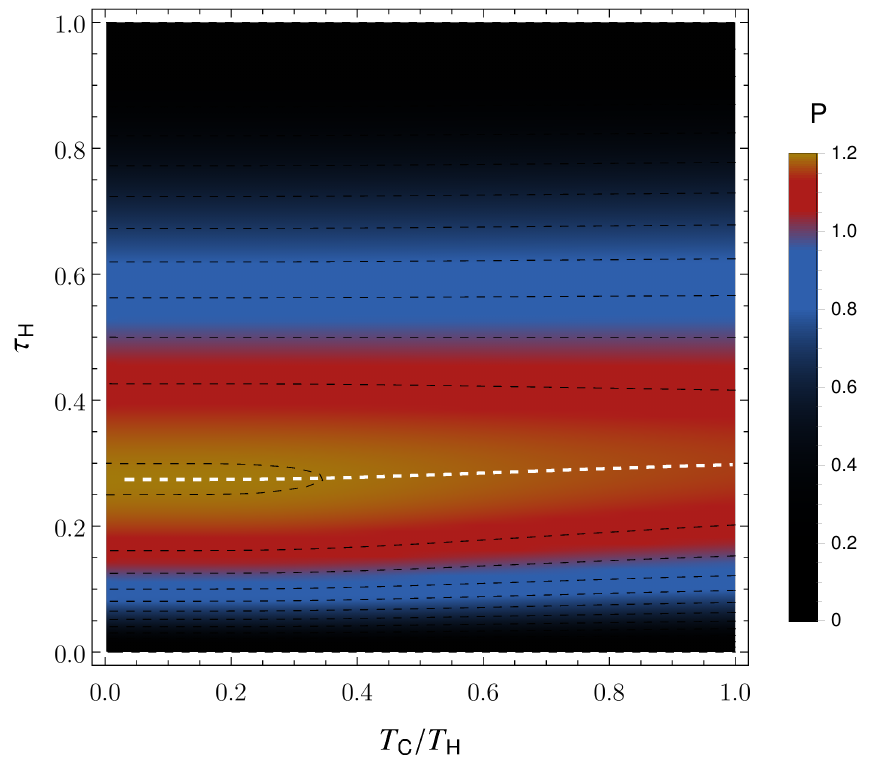}}
  \subfigure[$T\und{H}=3.0, \; \bar{\omega}\und{H} = 0.16$ and $\bar{\omega}\und{C}=0.69$.\label{densitiestratio_f}]{\includegraphics[angle = 0,width =3in]{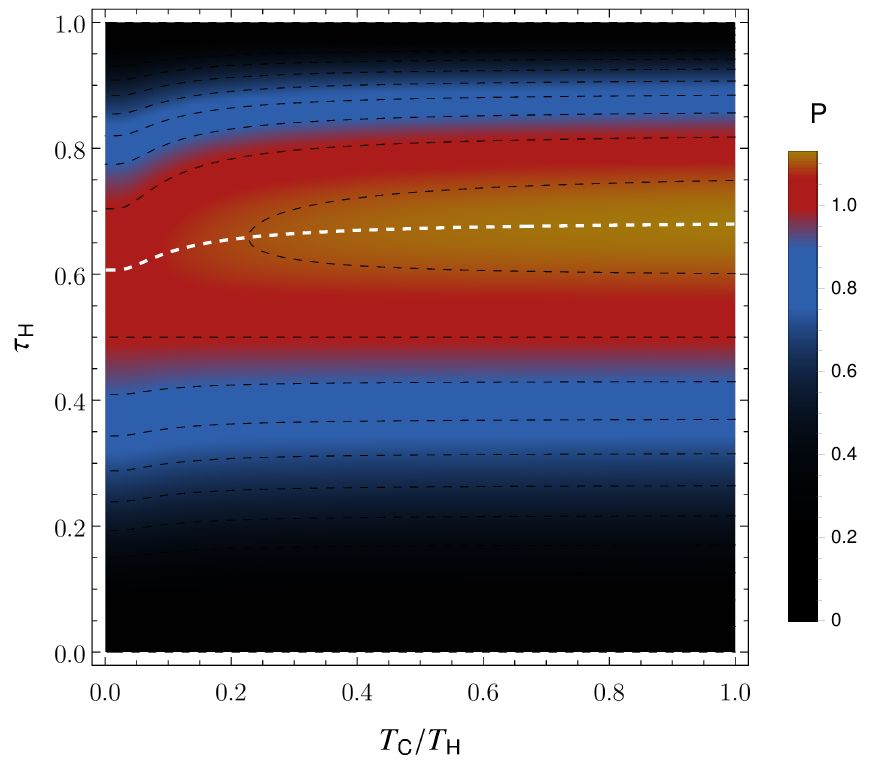}}\\
    \caption{Density plots of the power as a function of $\tau\und{H}$ and $T\und{C}/T\und{H}$ normalized by the power for $\tau\und{H} = \tau/2$. Each dashed black curve represents an increment of $0.1$ in the value of the power and the white dashed curve represents the value of $\tau\und{H}$ that maximizes the power for each value of $T\und{C} / T\und{H}$. For all panels, we set $\tau =1, \; \mu\und{H} = 0.4, \; \mu\und{C} = 0.6$ and $\epsilon = 1$.}
    \label{densitiestratio}
\end{figure*}

An interesting question that may be raised about this best protocol procedure is: how much do we gain by tuning $\tau\und{H}$ instead of just letting the QD be half of the time in contact with each reservoir (symmetric case)? In Fig.~\ref{densitiestratio}, we show the density plots of the ratio of the power output for $\tau\und{H}$ to the power output for the symmetric case  $\bar{P}(\tau\und{H})/\bar{P}(\tau/2)$ as a function of both $\tau\und{H}/\tau$ and $T\und{C}/T\und{H}$. As the temperature is changed (with  ${\bar \omega_{\rm i}}$'s, $\epsilon$ and the chemical potentials held fixed),  $t_\text{char}$ changes according to Eq. (\ref{temp}). Hence, comparisons between Figs.~\ref{densities} and~\ref{densitiestratio} should take this change into account.

Here, we used the same values of chemical potentials and energy as in Fig.~\ref{densities}. As can be seen, by properly tuning $\tau\und{H}$ one can increase the power output more than 25\% (Figs.~\ref{densitiestratio_a} and~\ref{densitiestratio_e}). However,  in order to the tuning  be more 
 effective, the transition rates $\bar{\omega}\und{H}$ and $\bar{\omega}\und{C}$ must be distinct, adding an asymmetry to the system --  Figs.~\ref{densitiestratio_c} and~\ref{densitiestratio_d} reveal that the ratio $\bar{P}(\tau\und{H})/\bar{P}(\tau/2)$ is less than or equal to 1 for $\bar{\omega}\und{H}=\bar{\omega}\und{C}$.
Finally, by comparing Fig.~\ref{densitiestratio_a} with Fig.~\ref{densitiestratio_b} and Fig.~\ref{densitiestratio_e} with Fig.~\ref{densitiestratio_f} we notice that the tuning of $\tau\und{H}$ can increase the power output either for $\bar{\omega}\und{H}$ greater or smaller than $\bar{\omega}\und{C}$, but it is for $\bar{\omega}\und{H} > \bar{\omega}\und{C}$ that the region of larger gain
corresponds to  the larger  efficiencies and power outputs (small $T\und{C}/T\und{H}$). 

\begin{figure}[ht]
\begin{center}
\includegraphics[width=.43\textwidth]{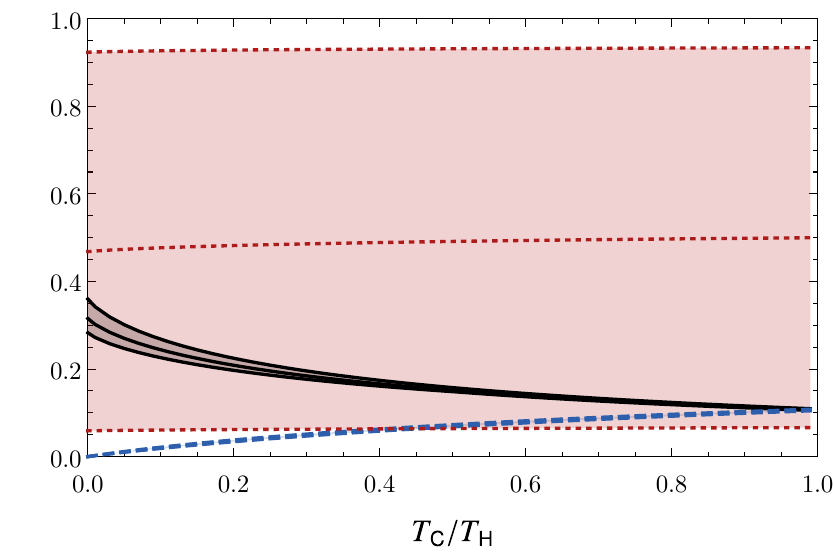}
\end{center}
\caption{Values of $\tau\und{H}/\tau$ (red dotted), $\omega\und{H}/\bar{\omega}\und{H}$ (black continuous) and $\omega\und{C}/\bar{\omega}\und{C}$ (blue dashed) after the maximization of power for fixed temperatures and $\tau=1$. The maximum, medium and minimum value for each quantity is presented and the highlighted areas between the three lines represents the possible values of each parameter after the optimization of the power for a grid of values of $\bar{\omega}\und{H}$ and $\bar{\omega}\und{C}$ ranging from 0.1 to 2 is steps of 0.05. }
\label{nunutau}
\end{figure}

\subsection{Complete maximization of power}

\begin{figure}[ht]
\begin{center}
\includegraphics[width=.43\textwidth]{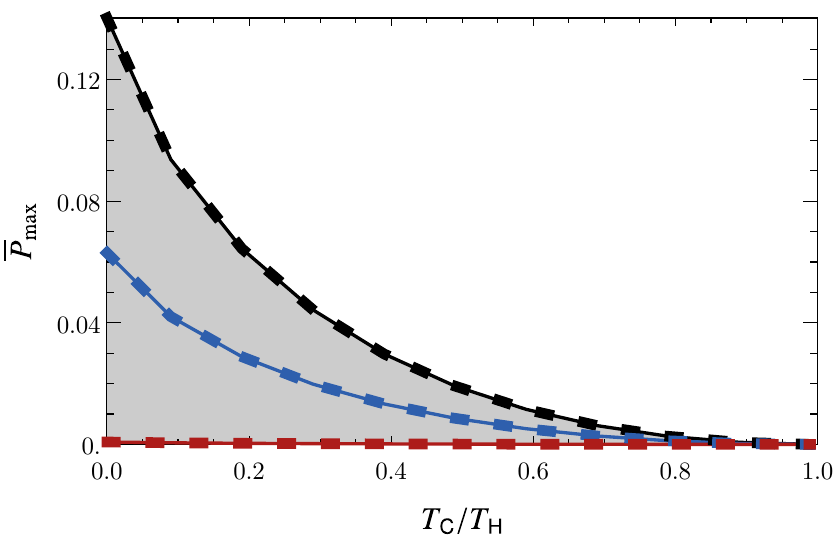}
\end{center}
\caption{Maximum  power output $\overline{P}_{\rm max}$ versus $T\und{C}/T\und{H}$ for $\tau=1$ and the same grid used in Fig.~\ref{nunutau}, where maximum (black), mean (blue) and minimum (red) values of maximum power are displayed. Continuous lines account for the maximization in terms of $\tau\und{H}$ only, while dashed lines represent the maximization in terms of $\tau\und{H}$, $\omega\und{H}/\overline{\omega}\und{H}$ and $\omega\und{C}/\overline{\omega}\und{C}$.} 
\label{pmax}
\end{figure}

In this section,  the complete maximization of the power,  not only
with respect to $\tau\und{H}$ (as in the previous section), but also taking 
into account  the transition rates is undertaken. Physically, this means that we are optimizing the physical properties (transition rates) as well as the machine protocol (time spent in contact with each reservoir).
Some remarkable features are depicted in Fig.~\ref{nunutau}, in which
the parameters $\tau\und{H}/\tau$, $\omega\und{H}$ and $\omega\und{C}$ that maximize the power output are considered as a function of the ratio $T\und{C}/T\und{H}$. That is, the optimization of the power output (with respect to $\tau\und{H}, \; \omega\und{H}$ and $\omega\und{C}$)
is undertaken  for all values of $\bar{\omega}\und{H}$ and $\bar{\omega}\und{C}$ in a square grid (from 0.1 to 2 in steps of 0.05). The
maximum, mean and minimum values of these quantities are then shown in Fig.~\ref{nunutau}.
Although the ratios $\omega\und{H}/\bar{\omega}\und{H}$
 and $\omega\und{C}/\bar{\omega}\und{C}$ vary slightly with the choice 
of the rates $\bar{\omega}\und{H}$
and $\bar{\omega}\und{C}$, an opposite trend is verified for $\tau\und{H}$,
 which is   very
 sensitive to the choice of transition rates/grids. This suggests that
  the choice of the time fraction in which the QD spends
 in contact with each reservoir is the  most important parameter
 for the power maximization.
 Such finding is reinforced by examining
   the behavior of output powers in Fig.~\ref{pmax} in which,
   in addition to the  optimal choices of $\omega\und{H}$ and $\omega\und{C}$,  one also takes  ``reliable'' estimates for them  and only the optimization
   with respect to the   $\tau\und{H}$ is carried out.
Although the power output strongly depends on the choices  of 
rates/grids $\bar{\omega}\und{H}$ and $\bar{\omega}\und{C}$ (see e.g.\ the lowest
and largest $\overline{P}_{\rm max}$ curves),  the difference between power outputs
coming from the optimization of
$\omega\und{i}$ and their mean values 
is very small
(not visible in the scale of the figure).
Finally, the difference between the lowest and largest $\overline{P}_{\rm max}$
is  due to its  monotonic increasing  by raising the transition rates.

\begin{figure}[ht]
\begin{center}
\includegraphics[width=.43\textwidth]{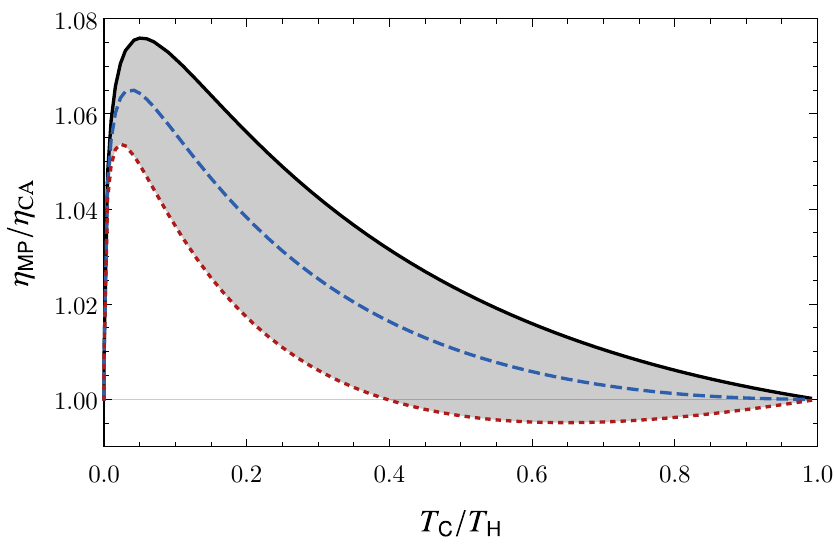}
\end{center}
\caption{Ratio between the efficiency at maximum power
$\eta\und{MP}$ and the Curzon-Ahlborn efficiency $\eta\und{CA}$
  versus $T\und{C}/T\und{H}$ for $\epsilon=1$ and $\tau=1$. The lines represent their locci of maximum (black continuous), mean (blue dashed) and minimum (red dotted) values obtained for the same grid used in Fig.~\ref{nunutau}.} 
\label{EMPECA}
\end{figure}

\begin{figure}[ht]
\begin{center}
\includegraphics[width=.43\textwidth]{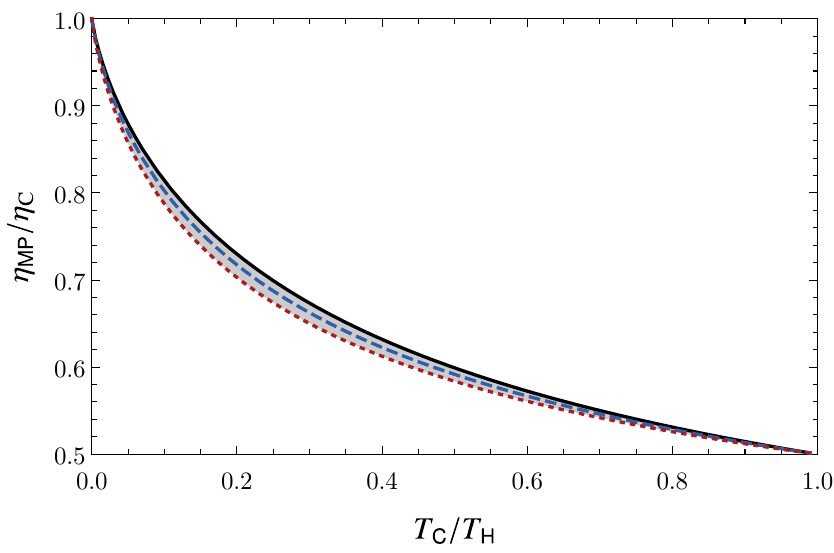}%
\end{center}
\caption{Ratio between efficiency at maximum power
$\eta\und{MP}$ and Carnot $\eta\und{C}$
  versus $T\und{C}/T\und{H}$ for $\epsilon=1$ and $\tau=1$. The lines represent their locci of maximum (black continuous), mean (blue dashed) and minimum (red dotted) values obtained for the same grid used in Fig.~\ref{nunutau}.}
\label{EMPEC}
\end{figure}

Lastly, the efficiency at maximal power $\eta\und{MP}$ is compared with the well-established Curzon-Ahlborn and Carnot ones in Figs.~\ref{EMPECA}
and~\ref{EMPEC}, respectively,
for different temperatures. Just as in Fig.~\ref{nunutau} and~\ref{pmax}, the optimization
is carried out for all values of $\bar{\omega}\und{H}$ and $\bar{\omega}\und{C}$ over a grid.
Except to some specific choices of $\overline{\omega}\und{H}$ and $\overline{\omega}\und{C}$ in a small range of $T\und{C}/T\und{H}$ between $0.4$ and $1$,
the maximization provides an efficiency slightly larger than the
Curzon-Ahlborn one. As for  Carnot and
Curzon-Ahlborn efficiencies,   Eqs.~(\ref{temp}) and~(\ref{eff}) show that
$\eta\und{MP}$
solely depends on the ratio of the temperatures and not on the specific value of $T\und{H}$.

Finally, $\eta\und{MP}$ is always (as it must be) 
  lower than $\eta_{\text{C}}$ (see e.g.\ Fig.~\ref{EMPEC}) for all set of
optimized parameters. Further, when $T\und{C}/T\und{H}\rightarrow$  $0$
and $1$,  all efficiencies collapse at the asymptotic values $1$ and $0$, respectively. 


\section{Discussion}\label{discuss}
 
In this contribution, we analyzed the role of asymmetric interaction times for optimizing the power output in thermal machines composed of  a quantum dot stochastic pump. Our findings showed that
a suited  power output optimization   can lead to a gain in the power larger than 25\% when compared
to the symmetrical case, in which the system is equally placed in contact with each reservoir.

A very remarkable point  regarding the importance
of present analysis concerns that the fine-tuning of the interaction time is expected
to be easier
to implement than an improved design of the machine itself.
Therefore, the present (exact) results may shed some light about the role of time protocol for the effectiveness of obtaining optimized power outputs.
  In fact, the choice of a ``good'' (instead of optimal) quantum dot
together with optimized time provides a reliable recipe for obtaining
an almost optimal (power output) machine, whose 
associate efficiencies are usually somewhat larger than the Curzon-Ahlborn one, but much larger than the symmetric case.

Further investigations of this model could explore other
regimes (e.g.\ as a refrigerator or a heater engine) and/or the modulation of the QD's energy to obtain mechanical work on top of the pumping of electrons.

\begin{acknowledgements}
PEH and CEF respectively acknowledge grant\#2017/24567-0 and grant\#2018/02405-1, S\~{a}o Paulo Research Foundation (FAPESP). AR thanks Pronex/Fapesq-PB/CNPq Grant No. 151/2018 and CNPq Grant No. 308344/2018-9. This study was financed in part by the Coordena\c{c}\~{a}o de Aperfei\c{c}oamento de Pessoal de N\'{i}vel Superior -- Brasil (CAPES) -- Finance Code 001.
\end{acknowledgements}

\bibliographystyle{apsrev4-1}
\bibliography{references}

\end{document}